\documentclass[journal,9pt]{IEEEtran}
\usepackage{amssymb}
\usepackage{amsmath}
\usepackage{amsfonts}
\usepackage{algorithm}
\usepackage{algorithmic}
\usepackage{graphicx}

\makeatletter

\newtheorem{theorem}{Theorem}

\newtheorem{problem}[theorem]{Problem}

\newtheorem{remark}[theorem]{Remark}

\numberwithin{equation}{section}
\numberwithin{equation}{subsection}
\numberwithin{theorem}{subsection}
\newcommand{\Z}{\mathbb{Z}}
\input{tcilatex}
\begin{document}

\title{Almost Linear Complexity Methods for Delay-Doppler Channel Estimation\smallskip 
}
\author{{\Large Alexander Fish and Shamgar Gurevich}
\thanks{%
This material is based upon work supported by the Defense Advanced Research
Projects Agency (DARPA) award number N66001-13-1-4052. 
This work was also supported in part by NSF Grant DMS-1101660 - "The
Heisenberg--Weil Symmetries, their Geometrization and
Applications".\smallskip {}}
\thanks{%
A. Fish is with School of Mathematics and Statistics, University of Sydney,
Sydney, NSW 2006, Australia. Email: alexander.fish@sydney.edu.au.} \thanks{%
S. Gurevich is with the Department of Mathematics, University of Wisconsin,
Madison, WI 53706, USA. Email: shamgar@math.wisc.edu.} }
\maketitle

\begin{abstract}
A fundamental task in wireless communication is \textit{channel estimation}:
Compute the channel parameters a signal undergoes while traveling from a
transmitter to a receiver. In the case of delay-Doppler channel, i.e., a
signal undergoes only delay and Doppler shifts, a widely used method to
compute  delay-Doppler parameters is the \textit{pseudo-random} method.
It uses a pseudo-random sequence of length $N,$ and, in case of non-trivial
relative velocity between transmitter and receiver, its computational
complexity is $O(N^{2}\log N)$ arithmetic operations. In \cite{FGHSS} the flag method was introduced to provide a faster algorithm for delay-Doppler channel estimation. It uses specially designed flag sequences and its complexity is $O(rN \log{N})$ for channels of \textit{sparsity} $r$.
In these notes, we introduce the \textit{incidence} and \textit{cross} methods for channel estimation. 
They use triple-chirp and double-chirp sequences of length $N$, correspondingly. These sequences are closely related to chirp sequences widely used in radar systems.
The arithmetic complexity of  the incidence and cross methods is $O(N\log{N}+r^3)$, and
 $O(N\log{N}+r^2)$, respectively. 
\end{abstract}



\section{\textbf{Introduction\label{In}}}

\PARstart{A}{ basic} building block in many wireless communication
protocols is \textit{channel estimation}: learning the channel parameters a
signal undergoes while traveling from a transmitter to a receiver \cite{TV}.
In these notes  we develop  efficient algorithms for delay-Doppler
(also called time-frequency) channel estimation. Our algorithms provide a
striking improvement over current methods in the presence of a substantial Doppler effect. 
Throughout these notes we denote by $\Z_N$ the set of integers $\{0,1,...,N-1\}$ equipped with addition and multiplication modulo $N$. We will assume, for simplicity, that $N$ is an odd prime. We denote by $\mathcal{H}=%
\mathbb{C}
(%
\mathbb{Z}
_{N})$ the vector space of complex valued functions on  $%
\mathbb{Z}
_{N}$, and refer to it as the \textit{Hilbert space of sequences}.  

\subsection{\textbf{Channel Model}}

We  describe the discrete channel model which was derived in \cite{FGHSS}.
 We assume that a transmitter uses a sequence $S \in \mathcal{H}$ to generate an analog waveform $S_A \in L^2(\mathbb{R})$ with bandwidth $W$ and a carrier frequency $f_c \gg W$. 
 Transmitting $S_A$, 
the receiver obtains the analog waveform $R_A \in L^2(\mathbb{R})$. We make the sparsity assumption on the number of paths for propagation of the waveform $S_A$. As a result, we have\footnote{%
In these notes $i$ denotes  $\sqrt{-1}.$}%
\begin{equation}
\label{cont_channel}
R_{A}(t)= \sum_{k=1}^{r}\beta _{k}\cdot \exp (2\pi if_{k}t)\cdot
S_{A}(t-t_{k})+\mathcal{W(}t), 
\end{equation}%
where $r$---called the \textit{sparsity }of the channel---denotes the number
of paths, $\beta _{k}\in 
\mathbb{C}
$ is the \textit{attenuation coefficient}, $f_{k}\in 
\mathbb{R}
$ is the \textit{Doppler shift} along the $k$-th path, $t_{k}\in 
\mathbb{R}
_{+}$ is the \textit{delay} associated with the $k$-th path, and $\mathcal{%
W}$ denotes a random white noise. We assume the normalization $%
\sum_{k=1}^{r}\left\vert \beta _{k}\right\vert ^{2}\leq 1$. The Doppler
shift depends on the relative velocity, and the delay encodes the distance along a path,
between the transmitter and the receiver. We will call 
\begin{equation}
(\beta_k, t_{k},f_{k}),\text{ }k=1,...,r,  \label{CP}
\end{equation}%
\textit{channel parameters,} and the main objective of channel detection is to estimate them.

\subsection{\textbf{Channel Estimation Problem}}

Sampling the waveform $R_A$ at the receiver side, with sampling rate $1/W$, we 
obtain a sequence $R \in \mathcal{H}$. It satisfies

\begin{equation}
\label{discr_channel}
R[n] = H(S)[n] + \mathcal{W}[n],
\end{equation}
where $H$, called  the \textit{channel operator},  acts  on $S \in \mathcal{H}$  by\footnote{We denote  $e(t) = \exp(2\pi i t/N)$.} 
\begin{equation}
\label{operator}
H(S)[n] = \sum_{k=1}^{r} \alpha_k e(\omega_k n) S[n - \tau_k], \mbox{  } n \in \Z_N, 
\end{equation}
with  $\alpha_k$'s are the complex-valued (digital) attenuation coefficients, $\sum_{k} | \alpha_k |^2 \leq 1$, $\tau_k \in \Z_N$ is the (digital) delay associated with the path $k$, $\omega_k \in \Z_N$ is the (digital) Doppler shift associated with path $k$, and $\mathcal{W}$ denotes the random white noise.
We will assume that all the coordinates of $\mathcal{W}$ are independent identically distributed random variables of  expectation zero.
\smallskip

\begin{remark}
\label{RDRD}The relation between the physical (\ref{CP}) and the discrete channel parameters is as follows (see Section I.A. in \cite{FGHSS}
and references therein): If a standard method suggested by sampling theorem
is used for the discretization, and $S_{A}$ has bandwidth $W$, then $\tau
_{k}=t_{k}W$ modulo $N$, and $\omega _{k}=Nf_{k}/W$ modulo $N$,
provided that $t_{k}\in \frac{1}{W}\mathbb{Z},$ and $f_{k}\in \frac{W}{N}%
\mathbb{Z},$ $k=1,...,r.$ In particular, we note that the integer $N$
determines the frequency resolution of the channel detection, i.e., the
resolution is of order $W/N.$

\end{remark}
\smallskip

The objective of  delay-Doppler channel estimation is: \smallskip

\begin{problem}[\textbf{Channel Estimation}]
\label{CE}Design $S\in \mathcal{H}$, and an effective method for extracting
the channel parameters $(\alpha _{k},\tau _{k},\omega _{k})$, $\ k=1,...,r,$
using $S$ and $R$ satisfying (\ref{discr_channel}).
\end{problem}


\begin{figure}[ht]
\includegraphics[clip,height=4cm]{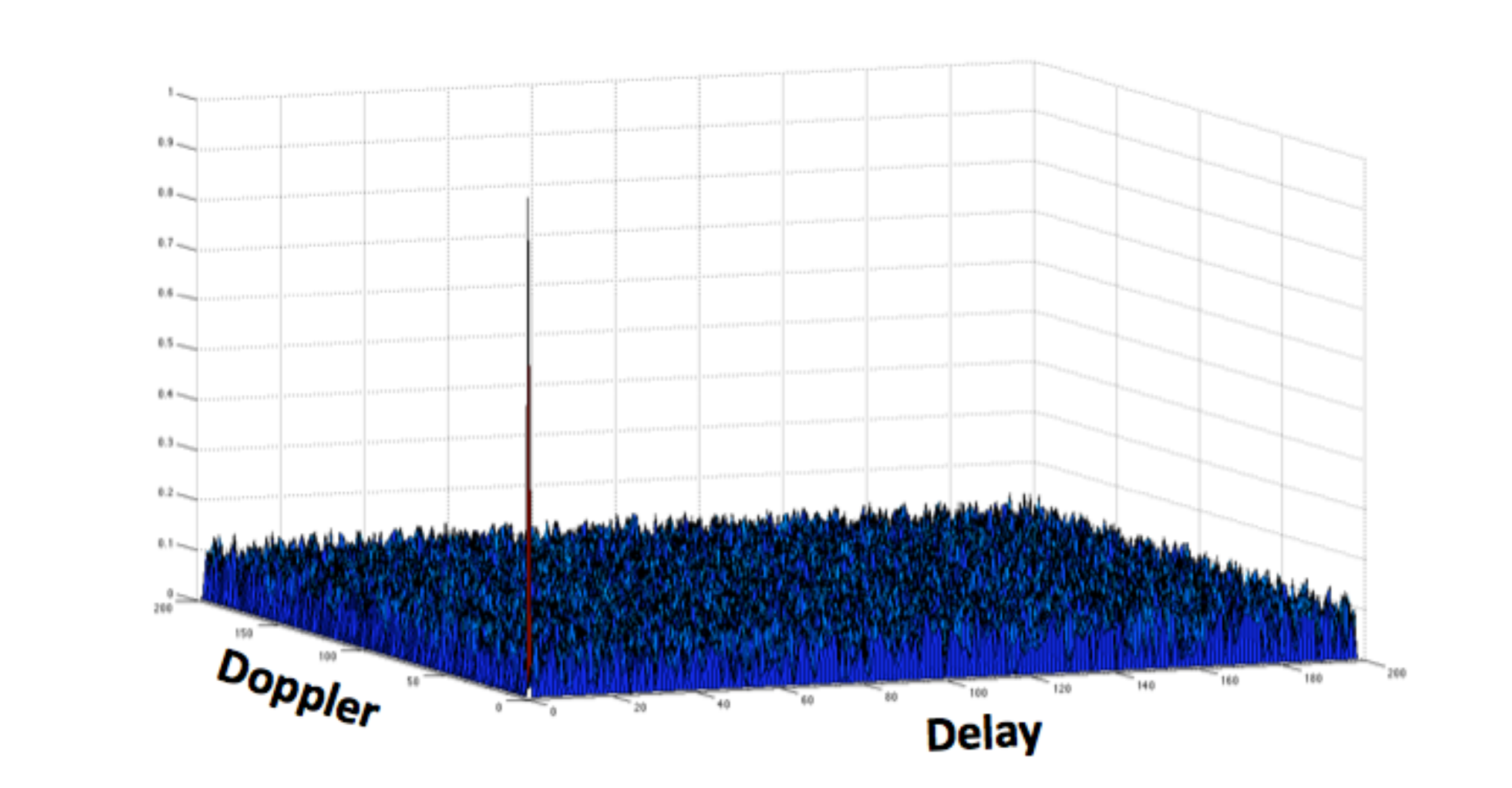}\\
\caption{Profile of $\mathcal{A}(\protect%
\varphi ,\protect\varphi )$ for $\protect\varphi $ pseudo-random sequence.}
\label{PRFigure}
\end{figure}

\subsection{\textbf{Ambiguity Function and Pseudo-Random Method}}

A classical method to estimate the channel parameters in (\ref{discr_channel}) is the 
\textit{pseudo-random method }\cite{GG, GHS, HCM, TV, V}. It uses two
ingredients - the ambiguity function, and a pseudo-random sequence.\smallskip

\subsubsection{\textbf{Ambiguity Function}}

In order to reduce the noise component in (\ref{discr_channel}), it is common to use the
ambiguity function that we are going to describe now. We
consider the \textit{Heisenberg operators }$\pi (\tau ,\omega ),$ $\tau
,\omega \in 
\mathbb{Z}
_{N},$ which act on $f\in \mathcal{H}$ by%
\begin{equation}
\left[ \pi (\tau ,\omega )f\right] [n]=e(-2^{-1}\tau \omega )\cdot e(\omega
n)\cdot f[n-\tau ],  \label{HO}
\end{equation}%
where $2^{-1}$ denotes $(N+1)/2,$ the inverse of $2$ $\func{mod}$ $N.$
Finally, the \textit{ambiguity function }of two sequences $f,g\in \mathcal{H}
$ is defined\footnote{%
For our purposes it will be convenient to use this definition of the
ambiguity function. The standard definition appearing in the literature is $%
A(f,g)[\tau ,\omega ]=\left\langle e(\omega n)f[n-\tau],g[n]\right\rangle .$%
} as the $N\times N$ matrix 
\begin{equation}
\mathcal{A(}f,g)[\tau ,\omega ]=\left\langle \pi (\tau ,\omega
)f,g\right\rangle ,\text{ \ }\tau ,\omega \in 
\mathbb{Z}
_{N},  \label{AF}
\end{equation}
where $\langle \mbox{ }, \mbox{ }\rangle$ denotes the standard inner product on $\mathcal{H}$.\smallskip
\begin{remark}[\textbf{Fast Computation of Ambiguity Function}]
\label{FC}The restriction of the ambiguity function to a line in the
delay-Doppler plane, can be computed in $O(N\log N)$ arithmetic operations
using fast Fourier transform \cite{R}. For more details, including explicit
formulas, see Section V of \cite{FGHSS}. Overall, we can compute the entire
ambiguity function in $O(N^{2}\log N)$ operations.\smallskip
\end{remark}

For $R$ and $S$ satisfying (\ref{discr_channel}), the law of the iterated logarithm
implies that, with probability going to one, as $N$ goes to infinity, we
have 
\begin{equation}
\mathcal{A}(S,R)[\tau ,\omega ]=\mathcal{A}(S,H(S))[\tau ,\omega
]+\varepsilon _{N},  \label{noise}
\end{equation}%
where $\left\vert \varepsilon _{N}\right\vert \leq \sqrt{2\log \log N}/\sqrt{%
N\cdot SNR},$ with $SNR$ denotes the \textit{signal-to-noise ratio}\footnote{%
We define $SNR=\left\langle S,S\right\rangle /\left\langle \mathcal{W},%
\mathcal{W}\right\rangle $.}.
\smallskip

\begin{remark}[Noise]
It follows from the equation (\ref{noise}), that for a reasonable noise level, it is sufficient to suggest a channel estimation method which finds channel parameters by analyzing the values of $\mathcal{A}(S,H(S))$.
\end{remark}

\subsubsection{\textbf{Pseudo-Random Sequences}}

\bigskip We will say that a norm-one sequence $\varphi \in \mathcal{H}$ is $%
B $-\textit{pseudo-random, }$B\in 
\mathbb{R}
$\textit{---}see Figure \ref{PRFigure} for illustration---if for every $\left(
\tau ,\omega \right) \neq (0,0)$ we have \textit{\ }%
\begin{equation}
\left\vert \mathcal{A}(\varphi ,\varphi )[\tau ,\omega ]\right\vert \leq B/%
\sqrt{N}. \label{pr}
\end{equation}%
There are several constructions of families of pseudo-random (PR) sequences
in the literature (see \cite{GG, GHS} and references therein).

\subsubsection{\textbf{Pseudo-Random Method}}

Consider a pseudo-random sequence $\varphi $, and assume for simplicity that 
$B=1$ in (\ref{pr}). Then we have 
\begin{eqnarray}
&&\mathcal{A}(\varphi ,H(\varphi ))[\tau ,\omega ]  \label{prm} \\
&=&\left\{ 
\begin{array}{c}
\widetilde{\alpha }_{k}+\tsum\limits_{j\neq k}\widetilde{\alpha }_{j}/\sqrt{N%
},\text{ \ if }\left( \tau ,\omega \right) =\left( \tau _{k},\omega
_{k}\right) ,\text{ }1\leq k\leq r; \\ 
\tsum\limits_{j}\widehat{\alpha }_{j}/\sqrt{N},\text{ \ \ \ \ \ otherwise, \
\ \ \ \ \ \ \ \ \ \ \ \ \ \ \ \ \ \ \ \ \ \ \ \ \ }%
\end{array}%
\right.  \notag
\end{eqnarray}%
where $\widetilde{\alpha }_{j},$ $\widehat{\alpha }_{j},$ $1\leq j\leq r,$
are certain multiples of the $\alpha _{j}$'s by complex numbers of absolute
value less or equal to one. In particular, we can compute the delay-Doppler
parameter $\left( \tau _{k},\omega _{k}\right) $ if the associated
attenuation coefficient $\alpha _{k}$ is sufficiently large. It appears as a peak
of $\mathcal{A}(\varphi ,H(\varphi )).$ Finding the peaks of $\mathcal{A}(\varphi,H(\varphi))$  constitutes the pseudo-random method.
Notice that the arithmetic complexity of the
pseudo-random method is $O(N^{2}\log N),$ using Remark \ref{FC}. 
For applications to sensing, that require sufficiently high frequency
resolution, we will need to use sequences of large length $N$. Hence, the following is a natural problem.
\smallskip

\begin{problem}[\textbf{Arithmetic Complexity}]
Solve Problem \ref{CE}, with method for extracting the channel parameters which requires almost linear arithmetic complexity.
\end{problem}
\smallskip

\begin{figure}[ht]
\includegraphics[clip,height=5cm]{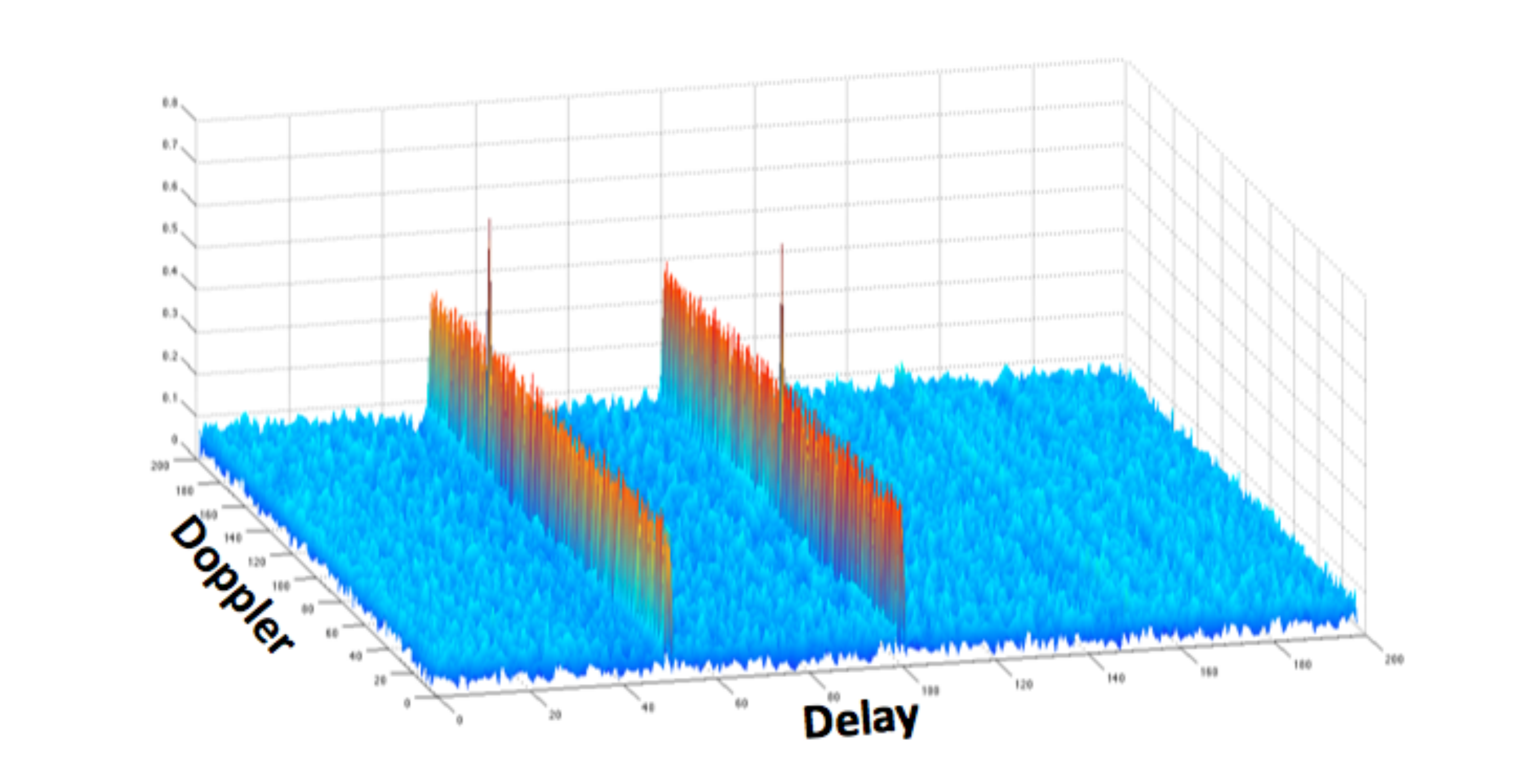}\\
\caption{Profile of $\mathcal{A(}%
f_{L},H(f_{L}))$ for flag $f_{L},$ $L=\{(0,\protect\omega )\},$ $N=199,$ and
channel parameters $(50,150)$, $(100,100),$  with attenuation
coefficients $0.7,$ $0.7,$ respectively.}
\label{FlagM}
\end{figure}

\subsection{\textbf{Flag Method}}
 In \cite{FGHSS} the flag
method was introduced in order to deal with the complexity problem. It
computes the $r$ channel parameters in $O(rN\log N)$ arithmetic operations. For
a given line $L$ in the plane $%
\mathbb{Z}
_{N}\times 
\mathbb{Z}
_{N},$ one construct a sequence $f_{L}$---called flag---with ambiguity
function $\mathcal{A(}f_{L},H(f_{L}))$ having special profile---see Figure %
\ref{FlagM} for illustration. It is essentially supported on shifted lines
parallel to $L$, that pass through the delay-Doppler shifts of $H$, and
have peaks there. This suggests a simple algorithm to extract the channel
parameters. First compute $\mathcal{A(}f_{L},H(f_{L}))$ on a line $M$
transversal to $L,$ and find the shifted lines on which $\mathcal{A(}%
f_{L},H(f_{L}))$ is supported. Then compute $\mathcal{A(}f_{L},H(f_{L}))$ on
each of the shifted lines and find the peaks. The overall complexity of the
flag algorithm is therefore $O(rN\log N)$, using Remark \ref{FC}. If $r$ is large, it might be computationally insufficient. 

\subsection{\textbf{Incidence and Cross Methods}}

In these notes we suggest two new schemes for channel estimation that  have much better arithmetic complexity than previously known methods. The schemes are based on the use of double and triple chirp sequences. 


\subsubsection{\textbf{Incidence Method}}

\begin{figure}[ht]
\includegraphics[clip,height=5cm]{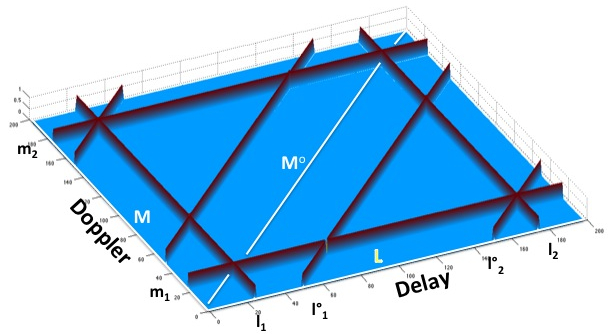}\\
\caption{Essential support of the ambiguity function $\mathcal{A}(C_{L,M,M^{\circ}}, H(C_{L,M,M^{\circ}}))$, where   $L$ is the delay line,  $M$ is the Doppler line, and $M^{\circ}$ is a diagonal line, and the support of $H$ consists two parameters. Points of $\Z_N \times \Z_N$ through them pass three lines are the true delay-Doppler parameters of $H$.}
\label{Incidence}
\end{figure}

We propose to use triple-chirp sequences for channel estimation. We associate with  three distinct lines $L,M$, and $M^{\circ}$ in $\Z_N \times \Z_N$, passing through the origin,  a sequence $C_{L,M,M^{\circ}} \in \mathcal{H}$. 
 This sequence has  ambiguity function essentially supported on the union of $L$, $M$, and $M^{\circ}$. As a consequence---see Figure \ref{Incidence} for illustration---the ambiguity function $\mathcal{A}(C_{L,M,M^{\circ}},H(C_{L,M,M^{\circ}}))$ is essentially 
 supported on the shifted lines $\{ (\tau_k,\omega_k) + ( L \cup M \cup M^{\circ}) \, | \, k=1,\ldots,r\}$. This observation, which constitutes the bulk of the incidence method,  enables a computation in  $O(N \log{N} + r^3)$ arithmetic operations  of all the time-frequency shifts  (see Section \ref{IncidenceMethod}). In addition, the estimation of the corresponding  $r$  attenuation
  coefficients takes 
 $O(r)$ operations. Hence, the overall complexity of incidence method is  $O(N \log{N} + r^3)$  operations.
 
\subsubsection{\textbf{Cross Method}}

\begin{figure}[ht]
\includegraphics[clip,height=4cm]{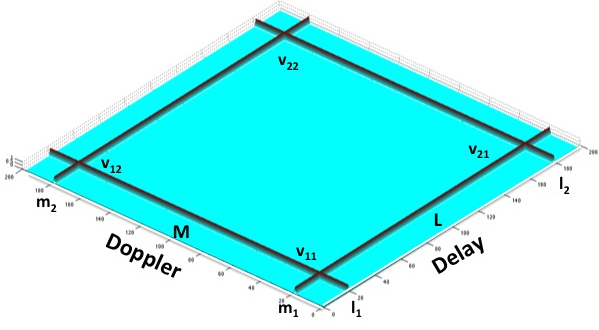}\\
\caption{Essential support of the ambiguity function $\mathcal{A}(C_{L,M}, H(C_{L,M}))$, where   $L$ is the delay line,  $M$ is the Doppler line, and the support of $H$ consists two parameters.}
\label{Cross}
\end{figure}

We propose to use double-chirp sequences for channel estimation. For two distinct lines $L$ and $M$ in $\Z_N \times \Z_N$, passing through the origin, we introduce a sequence $C_{L,M} \in \mathcal{H}$ with   ambiguity function supported on $L$, and  $ M$. 
Under genericity assumptions---see Figure \ref{Cross} for illustration---the essential support of $\mathcal{A}(C_{L,M},H(C_{L,M}))$ lies on $r \times r$ grid generated by shifts of the lines $L$, and $M$. Denote by $v_{ij} = l_i + m_j, \mbox{ }   l_i \in L, m_j \in M;\mbox{ }1 \le i,j \le r,$ the intersection points of the lines in the grid.
Using Remark \ref{FC} we  find all the points $v_{ij}, 1 \le i,j \le r,$ in $O(N \log{N})$ operations. The following matching problem arises: Find the $r$ points from $v_{ij}$, $1\leq i,j\leq r,$ which
belong to the support of $H$.
To suggest a solution, we use the values of the ambiguity function 
  to define  a certain simple hypothesis function $h: L \times M \to \mathbb{C}$ (see Section \ref{CrossMethod}). We obtain:
\begin{theorem}[\textbf{Matching}]
\label{TM}Suppose $v_{ij} = l_{i}+m_{j}$ is a delay-Doppler shift of $H,$ then $h(l_{i},m_{j})=0. $
\end{theorem}
\smallskip

\smallskip

The cross method makes use of Theorem \ref{TM} and checks the values  $h(l_i,m_j), \mbox{ }1 \le i,j \le r$. If a value is less than a priori chosen threshold, then the algorithm returns  $v_{ij} = l_i +m_j$ as one of the delay-Doppler parameters. To estimate the attenuation coefficient corresponding to $v_{ij}$
takes $O(1)$ arithmetic operations (see details in Section \ref{CrossMethod}).  
 Overall, the cross method enables  channel estimation in  $O(N \log{N} + r^2)$ arithmetic operations.

\section{\textbf{Chirp, Double-Chirp, and Triple-Chirp Sequences\label{CS} }}

In this section we introduce the chirp, double-chirp, and triple-chirp sequences, and discuss their correlation
properties. 

\subsection{\textbf{Definition of the Chirp Sequences}}

We have $N+1$ lines\footnote{%
In these notes by a \textit{line} $L\subset V$, we mean a line through $%
(0,0). $}
 in the \textit{discrete
delay-Doppler plane }$V=%
\mathbb{Z}
_{N}\times 
\mathbb{Z}
_{N}.$ For each $a\in 
\mathbb{Z}
_{N}$ we have the line $L_{a}=\left\{ (\tau ,a\tau );\tau \in 
\mathbb{Z}
_{N}\right\} $ of finite slope $a,$ and in addition we have the line of
infinite slope $L_{\infty }=\left\{ (0,\omega );\text{ }\omega \in 
\mathbb{Z}
_{N}\right\} .$ We have the orthonormal basis for $\mathcal{H}$
\begin{equation*}
\mathcal{B}_{L_{a}}=\left\{ C_{L_{a,b}}\text{; }b\text{ }\in 
\mathbb{Z}
_{N}\right\} ,
\end{equation*}%
 of \textit{%
chirp sequences} associated with $L_{a}$,
where 
\begin{equation*}
C_{L_{a,b}}[n]=e(2^{-1}an^{2}-bn)/\sqrt{N},n\in 
\mathbb{Z}
_{N}.
\end{equation*}%
In addition, we have the orthonormal basis 
\begin{equation*}
\mathcal{B}_{L_{\infty }}=\left\{ C_{L_{_{\infty },b}}\text{ ; }b\text{ }\in 
\mathbb{Z}
_{N}\right\} ,
\end{equation*}%
of chirp sequences associated
with $L_{\infty }$,
where 
\begin{equation*}
C_{L_{_{\infty },b}}=\delta _{b},
\end{equation*}%
denotes the Dirac delta sequence supported at $b.$  

\begin{figure}[ht]
\includegraphics[clip,height=4cm]{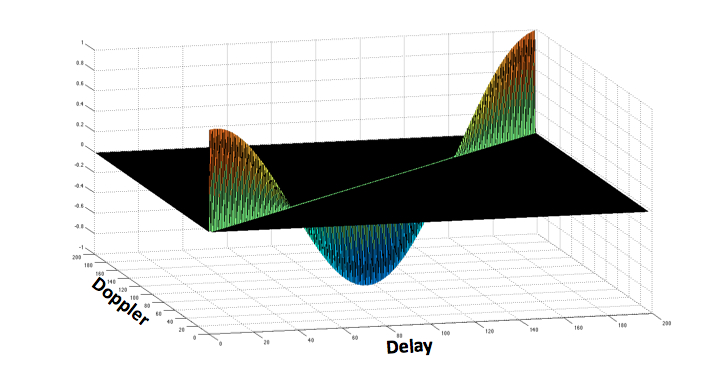}\\
\caption{Plot (real part) of $\mathcal{A}%
(C_{L_{1,1}},C_{L_{1,1}}),$ for chirp $C_{L_{1,1}}[n]=e[2^{-1}n^{2}-n]$,
associated with the line $L_{1}=\{(\protect\tau ,\protect\tau )\}.$ }
\label{ChirpD}
\end{figure}

\subsection{\textbf{Chirps as Eigenfunctions of Heisenberg Operators\label{E}%
}}

The Heisenberg operators (\ref{HO}) satisfy the commutation relations 
\begin{equation}
\pi (\tau ,\omega )\pi (\tau ^{\prime },\omega ^{\prime })=e(\omega \tau
^{\prime }-\tau \omega ^{\prime })\cdot \pi (\tau ^{\prime },\omega ^{\prime
})\pi (\tau ,\omega ),  \label{C}
\end{equation}%
for every $(\tau ,\omega ),(\tau ^{\prime },\omega ^{\prime })\in V.$ In
particular, for a given line $L\subset V,$ we have the family of commuting
operators $\pi (l),$ $l\in L.$ Hence they admit an orthonormal basis $\mathcal{B}_{L}$ for $%
\mathcal{H}$ of common eigenfunctions. Important property of the chirp
sequences is that for  every 
chirp sequence $C_L \in \mathcal{B}_L$, there exists a character\footnote{We denote by $\mathbb{C}^*$ the set of non-zero complex numbers} $\psi_L:L \to \mathbb{C}^*$, i.e. $\psi_L(l + l') = \psi_L(l) \psi_L(l'),$  $l,l' \in L$, such that 
\[
\pi_L(l) C_L = \psi_L(l) C_L, \mbox{ for every } l \in L.
\] 
This implies---see Figure \ref{ChirpD}---that for every $C_L \in \mathcal{B}_L$ 
we have

\begin{equation}
\label{chirpauto}
\mathcal{A}(C_{L},C_{L})[v]=\left\{ 
\begin{array}{c}
 \psi_L(v)\text{ \ if }v\in L; \\ 
\ \ 0\text{ \ \ \ \ \ \ \ if }v\notin L.\text{ \ }%
\end{array}%
\right.
\end{equation}%
It is not hard to see \cite{HCM} that for distinct lines $L$, and $M$, and two chirps $C_L \in \mathcal{B}_L, C_M \in \mathcal{B}_M$ we have

\begin{equation}
\label{AlmostOrthogonality}
\left\vert \mathcal{A(}C_{L},C_{M})[v]\right\vert =1/\sqrt{N}, \ \ \ \ \mbox{ for every } v \in V.
\end{equation}

\subsection{\textbf{Double-Chirp Sequences}}

For any two distinct lines $L,M \in V$, and two characters $\psi_L,\psi_M$ on them, respectively, denote by $C_L$ the chirp  corresponding to $L$ and $\psi_L$, and by $C_M$ the chirp  corresponding to $M$, and $\psi_M$.
We define  the \textit{double-chirp} sequence 
\[
C_{L,M} = (C_L + C_M)/\sqrt{2}.
\]
It follows from (\ref{chirpauto}) and (\ref{AlmostOrthogonality}) that for the line $K= L$, or $M$, we have
\[
\mathcal{A}(C_{K},C_{L,M})[v] \approx\left\{ 
\begin{array}{c}
 \psi_K(v)/\sqrt{2} \text{ \ \ \ if }v\in K; \\ 
\ \ 0\text{ \ \ \ \ \ \ \ \ \ \ \ \ \ \ \  if }v\notin K.\text{ \ }%
\end{array}%
\right.
\]

\subsection{\textbf{Triple-Chirp Sequences}}

Consider three distinct lines $L,M, M^{\circ} \in V$, and three characters $\psi_L,\psi_M, $ $\psi_{M^{\circ}}$ on them, respectively. Denote by $C_L, C_M$ and $C_{M^{\circ}}$ the chirps  corresponding to $L,M$ and $M^{\circ}$, and $\psi_L, \psi_M,$ and $\psi_{M^{\circ}}$,  
respectively.
We define  the \textit{triple-chirp} sequence 
\[
C_{L,M,M^{\circ}} = ( C_L + C_M + C_{M^{\circ}})/\sqrt{3}.
\]
It follows from (\ref{chirpauto}) and (\ref{AlmostOrthogonality}) that for the line $K=  L,M$ or $M^{\circ}$, we have
\[
\mathcal{A}(C_{K},C_{L,M,M^{\circ}})[v] \approx\left\{ 
\begin{array}{c}
 \psi_{K}(v)/\sqrt{3} \text{ \ \ \ \ if }v\in K; \\ 
\ \ 0\text{ \ \ \ \ \ \ \ \ \ \ \ \ \ \ \ \  if }v\notin K.\text{ \ }%
\end{array}%
\right.
\]

\section{\textbf{Incidence Method}}\label{IncidenceMethod}

We describe---see Figure \ref{Incidence} for illustration---the incidence algorithm.
\begin{algorithm}
\underline{\textbf{Incidence Algorithm}}\textbf{\bigskip }

\begin{description}
\item[\textbf{Input:}]$\,$Randomly chosen lines $L$, $M$, and $M^{\circ}$, and characters $\psi_L, \psi_M, \psi_{M^{\circ}}$ on them, respectively. 
Echo $R_{L, M, M^{\circ }}$ of the triple-chirp  $C_{L, M, M^{\circ }}$, threshold $T > 0$, and value of $SNR$.

\smallskip 

\item[\textbf{Output:}] $\,\,\,$ Channel parameters.

\end{description}

\smallskip

\begin{enumerate}

\item Compute $\mathcal{A(}C_{M},R_{L,M,M^{\circ}}\mathcal{)}$ on $L$, obtain peaks\footnote{We say that at  
$v \in V$ the ambiguity function of $C$ and $R$ has \textit{peak} if 
$|\mathcal{A}(C_L,R_L)[v]| \ge T \frac{\sqrt{2\log \log(N)}}{\sqrt{N \cdot SNR}}$} at $l_{1},...,l_{r_1}. \smallskip $

\item Compute $\mathcal{A(}C_{L},R_{L,M,M^{\circ}}\mathcal{)}$ on $M,$ obtain peaks at $m_{1},...,m_{r_2}.\smallskip $

\item Compute $\mathcal{A(}C_{M^{\circ }},R_{L,M,M^{\circ }}\mathcal{)}$ on $L,$
obtain peaks at $l_{1}^{\circ },...,l_{r_3}^{\circ }.\smallskip $

\item Find $v_{ij} = l_i + m_j$ which solve $l_{i}+m_{j}\in M^{\circ }+l_{k}^{\circ },$ $1\leq i\leq r_1$, $1\leq j\leq r_2$, $1\leq k\leq r_3$. \smallskip

\item For every delay-Doppler parameter $v_{ij} = l_i + m_j$ found in the previous step, compute $\alpha_{v_{ij}} =  \sqrt{3} \mathcal{A(}C_{L},R_{L,M,M^{\circ}}\mathcal{)}[m_j] \psi_L(l_i)$. Return the parameter $(\alpha_{v_{ij}},v_{ij})$.
\end{enumerate}
\end{algorithm}\footnotetext{%
We say that at $v\in V$ the ambiguity function of $f$ and $g$ has \textit{%
peak,} if $\left\vert \mathcal{A}(f,g)[v]\right\vert >T\sqrt{2\log \log N}/%
\sqrt{N\cdot SNR}.$}

\section{\textbf{Cross Method}}\label{CrossMethod}

Let  $C_{L,M}$ be the double-chirp sequence associated with the lines $L,M \subset V$, and  the characters $\psi_L$, and $\psi_M$, on $L$, and $M$, correspondingly. 
We define \textit{hypothesis} function $h:L\times M\rightarrow 
\mathbb{C}
$ by\smallskip 
\begin{eqnarray}
h(l,m) &=&\mathcal{A}(C_{L},R_{L,M})[m]\cdot \psi _{L}[l]  \label{h} \\
&&-\mathcal{A}(C_{M},R_{L,M})[l]\cdot e(\Omega \lbrack l,m])\cdot \psi
_{M}[m],  \notag
\end{eqnarray}%
where\footnote{%
In linear algebra $\Omega $ is called \textit{symplectic form.}} $\Omega
:V\times V\rightarrow 
\mathbb{Z}
_{N}$ is given by $\Omega \lbrack (\tau ,\omega ),(\tau ^{\prime },\omega
^{\prime })]=\tau \omega ^{\prime }-\omega \tau ^{\prime }.\smallskip $
Below we describe---see Figure \ref{Cross}---the Cross Algorithm.

\begin{algorithm}
\underline{\textbf{Cross Algorithm}}\smallskip 

\begin{description}

\item[\textbf{Input:}] $\,$
Randomly chosen lines $L$, $M$, and characters $\psi_L, \psi_M$ on them, respectively. Echo  $R_{L, M}$ of the double-chirp  $C_{L, M}$; thresholds $T_1, T_2 > 0$, and the value of $SNR$.

\smallskip 

\item[\textbf{Output:}] $\,\,\,$ Channel parameters.

\end{description}

\smallskip

\begin{enumerate}

\item Compute $\mathcal{A}(C_{M},R_{L,M})$ on $L,$ and take the $r_1$
peaks\footnote{We say that at $v \in V$ the ambiguity function of $f$ and $g$ has peak, if 
$\mathcal{A}(f,g)[v] \geq T \frac{\sqrt{\log \log (N)}}{\sqrt{N \cdot SNR}}$} located at points $l_{i},$ $1\leq i\leq r_1 \smallskip$.

\item \smallskip Compute $\mathcal{A}(C_{L},R_{L,M})$ on $M,$ and take the $r_2$
peaks located at the points $m_j, 1 \leq j \leq r_2 \smallskip$.

\item Find $v_{ij} = l_i +m_j$ which solve  $|h(l_{i},m_{j})| \leq T_2 \sqrt{2\log \log (N)}/\sqrt{N \cdot SNR}$, where $1\leq i \leq r_1$,  
$1\leq j\leq r_2$.\smallskip

\item For every delay-Doppler parameter $v_{ij} = l_i + m_j$ found in the previous step, compute $\alpha_{v_{ij}} =  \sqrt{2} \mathcal{A(}C_{L},R_{L,M}\mathcal{)}[m_j] \psi_L(l_i)$. Return the parameter $(\alpha_{v_{ij}},v_{ij})$.

\end{enumerate}
\end{algorithm}\footnotetext{%
We say that at $v\in V$ the ambiguity function of $f$ and $g$ has \textit{%
peak}, if
\par
$\left\vert \mathcal{A}(f,g)[v]\right\vert >T_{1}\sqrt{2\log \log N}/\sqrt{%
N\cdot SNR}.$}

\section{\textbf{Conclusions\label{Co}}}

\begin{figure}[ht]
\includegraphics[clip,height=3.5cm]{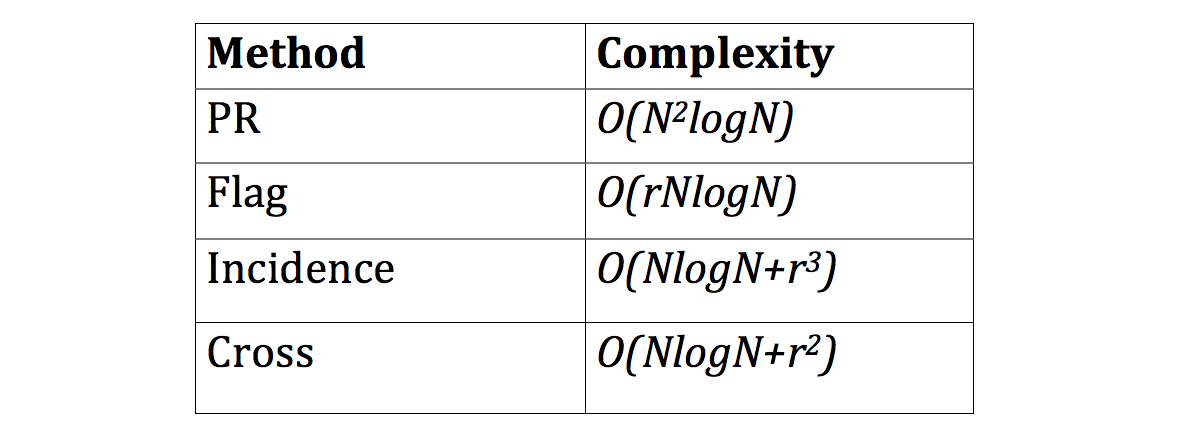}\\
\caption{Comparing methods, with respect to arithmetic complexity for $r$
channel parameters, and sensibility of the parameters in terms of magnitude of attenuation
coefficients (noiseless scenario). }
\label{Table}
\end{figure}

In these notes we present the incidence and cross methods for efficient
channel estimation. These methods, in particular, suggest solutions to  the arithmetic complexity problem. Low arithmetic complexity
enables  working with sequences of larger length $N$, and hence higher velocity resolution of channel parameters. We summarize these
important features in Figure \ref{Table}, and putting them in comparison
with the pseudo-random (PR) and Flag methods.

\vspace{-0.1in}

\begin{remark}
Both new methods are robust to a certain degree of noise since they use the values of the ambiguity functions, which is a sort of averaging.
\end{remark}
\smallskip

\textbf{Acknowledgements. }We are grateful to our collaborators A. Sayeed,
and O. Schwartz, for many discussions related to the research reported in
these notes.

\end{document}